\documentclass[conference]{IEEEtran}
\IEEEoverridecommandlockouts
\usepackage{cite}
\usepackage{amsmath,amssymb,amsfonts}
\usepackage{algorithmic}
\usepackage{graphicx}
\usepackage{textcomp}

\usepackage{xcolor}
\usepackage{listings}
\definecolor{codegreen}{rgb}{0,0.6,0}
\definecolor{codegray}{rgb}{0.5,0.5,0.5}
\definecolor{codepurple}{rgb}{0.58,0,0.82}
\definecolor{backcolour}{rgb}{0.95,0.95,0.92}
\lstdefinestyle{mystyle}{
    backgroundcolor=\color{backcolour},   
    commentstyle=\color{codegreen},
    keywordstyle=\color{magenta},
    numberstyle=\tiny\color{codegray},
    stringstyle=\color{codepurple},
    basicstyle=\ttfamily\footnotesize,
    breakatwhitespace=false,         
    breaklines=true,                 
    captionpos=b,                    
    keepspaces=true,                 
    numbers=left,                    
    numbersep=5pt,                  
    showspaces=false,                
    showstringspaces=false,
    showtabs=false,                  
    tabsize=2
}
\usepackage{hyperref}
\usepackage{xcolor}
\def\BibTeX{{\rm B\kern-.05em{\sc i\kern-.025em b}\kern-.08em
    T\kern-.1667em\lower.7ex\hbox{E}\kern-.125emX}}
\begin{document}

\title{Data Integration Framework for Virtual Reality Enabled Digital Twins 
}

\author{
\IEEEauthorblockN{Florian Stadtmann}
\IEEEauthorblockA{\textit{Department of Engineering Cybernetics} \\
\textit{NTNU}
Trondheim, Norway \\
florian.stadtmann@ntnu.no}
\and
\IEEEauthorblockN{Hary Pirajan Mahalingam}
\IEEEauthorblockA{\textit{Department of Computer Science} \\
\textit{NTNU}
Trondheim, Norway \\
harypm@stud.ntnu.no}
\and
\IEEEauthorblockN{Adil Rasheed}
\IEEEauthorblockA{\textit{Department of Engineering Cybernetics} \\
\textit{NTNU}
Trondheim, Norway \\
adil.rasheed@ntnu.no}  
}
\maketitle

\begin{abstract}
Digital twins are becoming increasingly popular across many industries for real-time data streaming, processing, and visualization. They allow stakeholders to monitor, diagnose, and optimize assets. Emerging technologies used for immersive visualization, such as virtual reality, open many new possibilities for intuitive access and monitoring of remote assets through digital twins. This is specifically relevant for floating wind farms, where access is often limited. However, the integration of data from multiple sources and access through different devices including virtual reality headsets can be challenging. In this work, a data integration framework for static and real-time data from various sources on the assets and their environment is presented that allows collecting and processing of data in Python and deploying the data in real-time through Unity on different devices, including virtual reality headsets. The integration of data from terrain, weather, and asset geometry is explained in detail. A real-time data stream from the asset to the clients is implemented and reviewed, and instructions are given on the code required to connect Python scripts to any Unity application across devices. The data integration framework is implemented for a digital twin of a floating wind turbine and an onshore wind farm, and the potential for future research is discussed.
\end{abstract}

\begin{IEEEkeywords}
data integration framework, digital twin, wind energy
\end{IEEEkeywords}

\section{Introduction} 
    
    Digital twins have become an important tool in many industries including, but not limited to, health, meteorology, manufacturing, education, and transportation \cite{Rasheed2020dtv}.
    A digital twin is often described as a virtual, digital, or software representation of a (physical) asset, object, thing, process, system, organization, person, or other abstraction (e.g. ~\cite{Gartnerdod, Martin2021wia, IMB2020wia, DNVGL2020dra, Parriswia, Siemensdts, Olcott2020dtc, dti, Rasheed2020dtv}).
    Depending on the exact definition, different additional capabilities are suggested or required from a digital twin.
    In many definitions, the digital twin is synchronized with the physical system (e.g. in~\cite{Martin2021wia, IMB2020wia, Parriswia, Oracleati, Olcott2020dtc, Rasheed2020dtv}). Some definitions suggest using digital twins to simulate, predict, optimize, and/or support decision-making (such as in ~\cite{IMB2020wia, Parriswia, Siemensdts, Rasheed2020dtv}). Finally, there are definitions that require a bidirectional data exchange~\cite{Trauer2020wia} between the physical system and its digital twin.
    
    A tool to successfully distinguish between digital twins with such varying capabilities, the digital twin capability level scale, is used in~\cite{Stadtmann2023dti, Sundby2021gcd, Stadtmann2023doa, DNVGL2020dra, Elfarri2023aid, Stadtmann2023sda}.
    According to the scale, a digital twin can be classified on a scale from zero to five. A standalone digital twin (level 0) is a virtual representation of an asset/system/process that is not synchronized with its physical counterpart. The descriptive digital twin (level 1) is updated with real-time data to describe the current asset/system/process state. At the diagnostic level (level 2), the synchronized virtual asset state is used for condition monitoring and fault detection. The predictive (level 3) digital twin predicts the future asset state. A prescriptive digital twin (level 4) uses uncertainty quantification and risk analysis to provide recommendations to the user. Finally, the autonomous digital twin (level 5) possesses automated bidirectional data exchange to control the asset.
    
    From the definitions and the capability level scale, it immediately becomes evident that a digital twin requires information on the asset and its environment. For capability levels 1-5, this information needs to be updated and processed in real-time. 
    However, a single data source is rarely sufficient to build a holistic up-to-date description of the asset and its environment. For different data types, sources, and formats, data integration can quickly become messy. 
    Furthermore, the digital twin may have to be accessed by multiple users from different devices. These devices are not only limited to desktop computers, laptops, tablets, and mobile phones. Virtual reality headsets are becoming more and more popular not only in the gaming industry but also in many engineering and management applications. The value of virtual reality technology in digital twins has been demonstrated for example in~\cite{Stadtmann2023doa,Elfarri2023aid,Stadtmann2023sda}.
    
    Wind energy is one of the application areas where the application of digital twins has a large potential, but is also especially challenging, as has been extensively covered in \cite{Stadtmann2023dti,Xia2023oam,Ciuriuc2022dtf}. 
    Recently, a digital twin of a wind farm using augmented reality has been demonstrated in \cite{Haghshenas2023pdt}, and as part of the work, a communication architecture is presented. In \cite{Issa2023dto} focus is put on integrating a digital twin of a simulated wind turbine into the Microsoft Azure IoT Platform.
    While both works provide valuable insights into data integration, they do not contain any real-time data from an actual full-scale wind turbine or its environment.
    \cite{Liu2023adt} proposes a framework for data integration of structure-related parameters and covers modeling strategies, but the data transfer is not addressed. 
    In \cite{Stadtmann2023doa} and \cite{Stadtmann2023sda}, the authors of this work have demonstrated a standalone, descriptive, and predictive digital twin of the onshore wind farm Bessakerfjellet and the floating offshore wind turbine Zefyros. 
    However, in both works, focus has been put on the predictive capabilities and on the 3D visualization. In contrast, the underlying data integration has barely been mentioned.

    This work presents a data integration framework, which has been successfully implemented into a digital twin of a commercial onshore wind farm and a digital twin of a full-scale floating wind turbine using actual data in real-time.
    Data from the floating wind turbine is streamed through the digital twin to the clients. The data accumulation across sources is performed by a Python server. A web-based real-time interface between the Python server and the clients has been established. Multiple clients are supported, and the digital twin can be controlled from a mobile phone.
    The data integration framework is explained in detail and gives recommendations on how to integrate static and dynamic asset data as well as environmental data for digital twins of wind farms and other assets.

    \begin{figure*}[h]
        \centering
        \includegraphics[width=\linewidth]{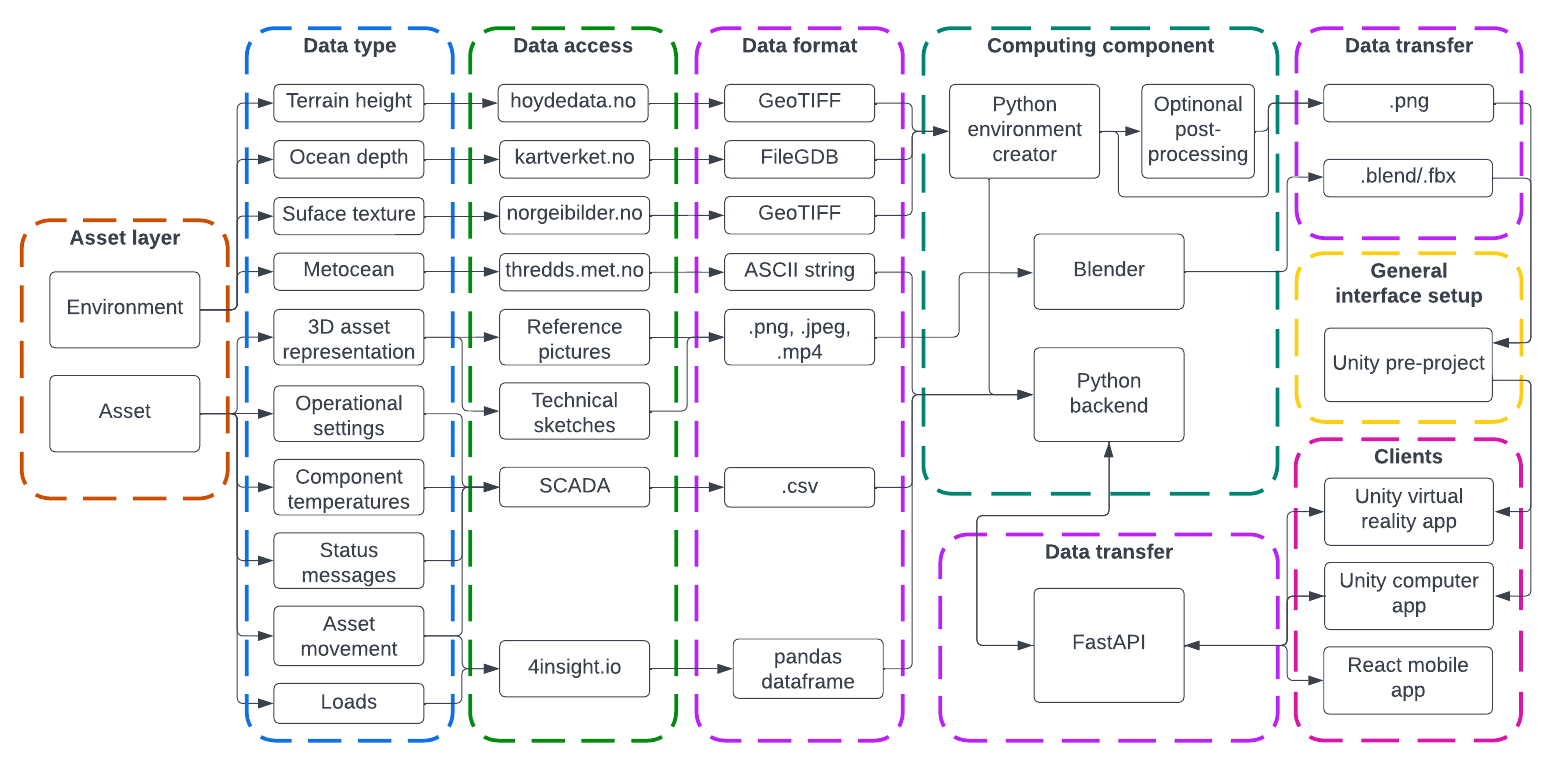}
        \caption{Data integration framework}
        \label{fig:framework}
    \end{figure*}

\section{Methodology and Implementation}
    In this section, the data integration framework and its implementation are presented for each of the components. The full framework is shown in Figure~\ref{fig:framework}. The first two subsections explain the clients and the computing component, whereafter the data stream of static and real-time data is explained.

    \subsection{Clients}
    A key feature of the digital twin is the human-machine interface, where users can interact with the digital twin. 
    To ensure scalability, multiple users need to be able to interact with the digital twin simultaneously.
    Furthermore, different users need to access the digital twin from different devices, including desktop computers, notebooks, and mobile phones with different operating systems. Deployment as an app increases the possible features and capabilities, while a web application is accessible from anywhere.
    Finally, it is important to note that users may not always possess technical expertise or domain knowledge. Therefore, the user interface should be designed as intuitively as possible. Virtual and augmented reality enable many new ways of presenting information intuitively, but the integration of data can be more challenging.
    
    In this work, the game engine Unity has been chosen as the software for the user clients.
    Unity allows building clients for desktop computers (Windows, Linux, Mac), mobile phones (Android, iOS), virtual reality headsets (Android), and 3D web applications (WebGL).
    Using one software for different types of clients reduces redundancy in the implementation of digital twins. Nonetheless, multiple different engines can be interfaced with the digital twin. Here, Unity is used for deployment on computers and virtual reality headsets, while RedactJS is used to build a dashboard for mobile phones. It should be noted that deployment of the app on virtual reality headsets limits the usability of existing data transfer libraries which requires additional effort as is explained in Section~\ref{ssec:frontback}.
    
    \subsection{Computing component}
    Performing computations on the client devices is disfavoured due to limited computation power, increased power consumption, and redundancy in computation when multiple clients exist. Furthermore, client devices are not constantly online/operating, making it impossible to continuously monitor the alarms. For these reasons, all computing tasks are performed outside of the client in a separate computing component.
    All data flows through the computing component, and relevant data is processed and combined for applications such as fault detection, virtual sensing, simulations, forecasts, uncertainty estimation, and what-if scenario analysis.
    Note that while it is presented here as a single component, it does not have to be hosted by a single machine or coded in any single programming language. It can be expected that, as the digital twin grows, there will be several modules in different languages running in parallel. Cloud computing may be used for heavy computing tasks, but due to the large amount of real-time data gathered, edge computing might be preferred over cloud computing. However, for many applications, even a consumer computer can already perform the necessary calculations. 
    
    In both implementations of this work, a consumer laptop (Razer Blade 15) has been used to represent the computing unit.
    The usage of various packages in different programming languages will require additional data interfaces.
    While it is possible to further distribute the computing tasks in this way, using a single programming language does not restrict the scalability. 
    Here, the programming language Python was used for all data import and computing. Python has become very popular among many industries and has extensive support through various packages, which makes it possible to import and process diverse data types from several sources.
    While Python has proven to be more than sufficient for this work, note that for tasks that are computationally expensive and time-sensible, there might be other programming languages that are better suited.

    \subsection{Terrain}
    Static and slow-changing data such as terrain data and general asset geometry are processed offline. 
    While a lot of terrain data is openly accessible, measurement campaigns are often only taken once every few years in irregular intervals, and online updating of this information is currently not required. Nonetheless, the data processing is automated where possible, and implemented in a way that would allow infusion of new information online.
    The offline data are combined to build a visual representation of the asset and its environment, as well as a static asset information model.

    For assets in an isolated environment, it may be sufficient to model its immediate environment manually. For assets or systems that are deployed/operated outdoors, the terrain (including infrastructure) becomes an important factor, and depending on size and required granularity, manual modeling can quickly become very tedious. Wind farms are but one example where terrain shape and land cover play an important role. Therefore, in the integration framework, a terrain processing pipeline is included, which processes the terrain offline, but can be easily reapplied when new environments need to be modeled for the same or other digital twins.
    
    For this implementation, terrain height measured through LIDAR is requested from the website Hoydedata~\cite{hoydedata} in GeoTIFF format. Ocean depth measured through sonar is downloaded through Geonorge~\cite{kartverket} as FileGDB (contour lines), and aerial pictures are extracted from Norge i bilder~\cite{norgeibilder} in GeoTIFF. The three data sets are loaded into Python by using the rasterio package for the terrain height and aerial pictures in GeoTIFF format and the geopandas and fiona packages for ocean contour lines. Once all data is converted into array shapes, the ocean coverage is identified from the terrain height, and its depth is calculated by interpolating the ocean contour lines to the relevant grid size using a k-d-tree for k-nearest-neighbor interpolation. Next ocean depth and heightmaps are combined into a single array and sliced and the aerial images are sliced into tiles with the same size and coordinates. On one hand, the slicing benefits scalability. On the other hand, it increases computational efficiency on the client side. The resulting tiles are normalized and saved in .png format, and metadata such as normalization factor and coordinates are saved as .txt. From there, the tiles are imported by a C\# script into the client, which is built using the Unity game engine. There finally a tiled 3D terrain is built by combining terrain shape with aerial pictures and meta-data. 
    
    Most of the process is automated, except for the initial data request from the respective websites. 
    Note also that the data sources cover the Norwegian land, coast, and sea, but the process can be directly reproduced for any data source that provides terrain height or ocean depth in the corresponding format. For other data formats, only minor changes have to be made. 

    \subsection{3D model of the asset}
    In addition to the environment, a 3D model of the asset is required. For mechanical assets, a 3D model has often already been created during the design stage. However, depending on the user case, these models are highly confidential.
    Photogrammetry provides a way to extract 3D models from pictures and is especially beneficial for applications where the asset shape is too large or complex to model by hand, such as for cities, or changes over time and needs continuous remodeling, as is the case for plants.
    For a moving asset with a known 3D shape, the position, and rotation can be tracked as demonstrated in~\cite{Sundby2021gcd} where motion detection, object detection, and pose estimation have been combined to track objects in an environment.

    Unfortunately, for most wind turbines, access to 3D models is heavily restricted. While photogrammetry has big potential, for the current work, a scanning campaign was not feasible due to a lack of access to the turbine. Therefore, two wind turbine models have been built manually in \cite{Stadtmann2023doa, Stadtmann2023sda} which have been improved here through a diverse set of pictures, videos, and technical sketches of the turbine and its components using the software Blender. The model can be saved in .blend or .fbx file format.

\begin{figure}[h]
        \centering
        \includegraphics[width=\linewidth]{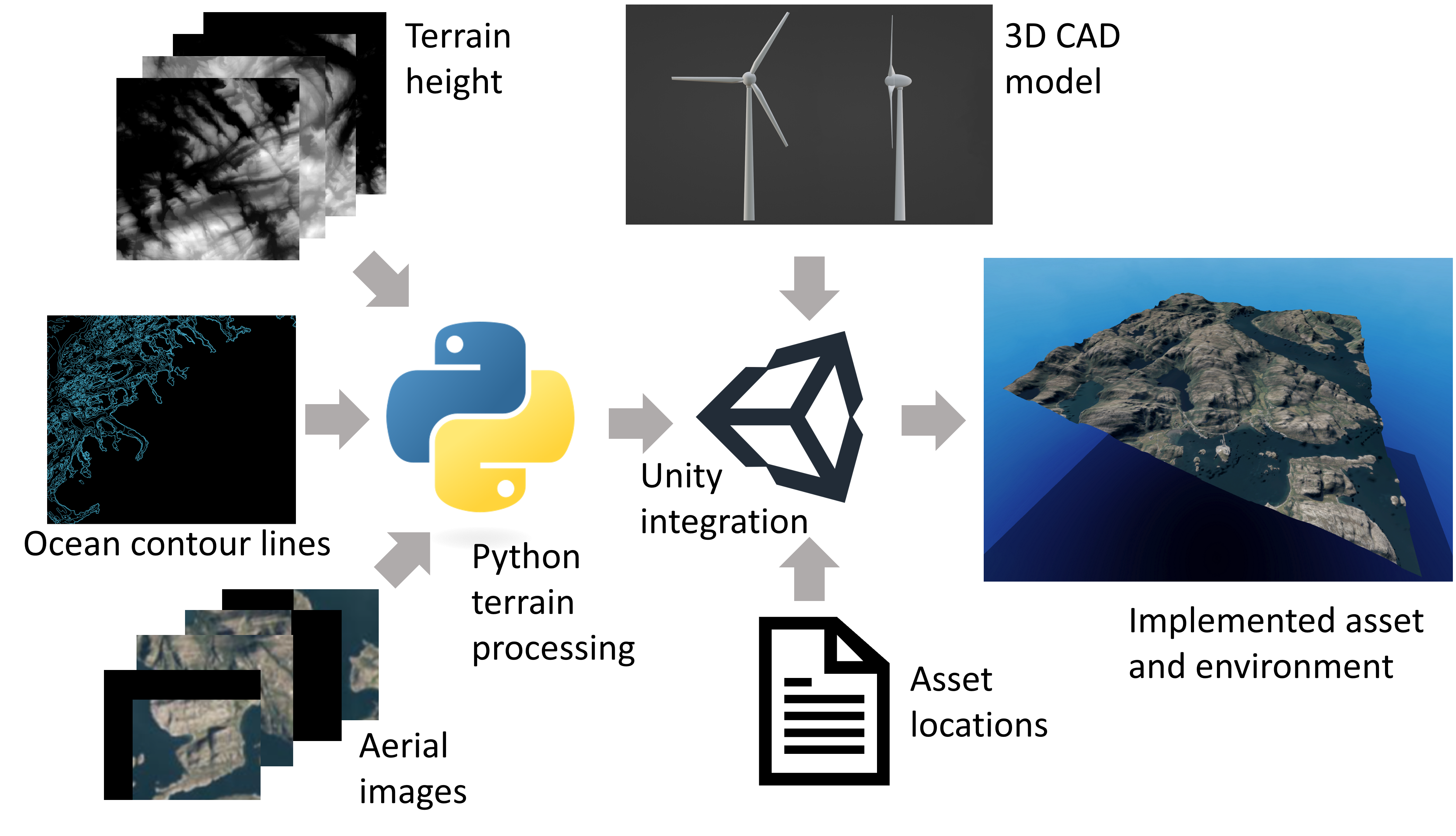}
        \caption{Implementation of static data}
        \label{fig:static}
    \end{figure}

    The finished 3D model is then imported and placed into the Unity environment through C\# scripts using a .txt file with corresponding coordinates. The same pipeline can be used for multi-object assets. Furthermore, components of the asset can be modified or replaced inside the Unity environment without the need for remodeling the whole asset. The full integration pipeline of the static data consisting of terrain and turbines is shown in Figure~\ref{fig:static}.

    \subsection{Metocean data}
    Changing meteorological and/or ocean conditions are highly relevant not only to outdoor assets but also to indoor assets where they can affect temperature and humidity. Even systems may be heavily impacted by weather conditions. For example, extreme weather such as storms or snow may affect transportation schedules, and rain and sunshine can impact mood and affect social systems. For wind energy, wind speed obviously plays an important role, but production can also be affected by wind direction,  pressure, temperature, humidity, and - for offshore systems - waves. Furthermore, precipitation impacts the turbine condition and can have severe effects on the health of the blades.
    
    Meteorological data is typically available through government-funded meteorological institutes. 
    In the Scandinavian area, the Norwegian Meteorological Institute provides nowcasts and forecasts as part of the MetCoOp ensemble system.
    The forecasts have a spatial resolution of $2.5~\mathrm{km}\times2.5~\mathrm{km}$, a temporal resolution of 1~hour, and a forecasting horizon of 61~hours. They contain a wide range of parameters including, but not limited to, wind speed, wind direction, pressure, temperature, humidity, precipitation, snowfall, and cloud coverage. More information about the underlying MEPS model is available at~\cite{Rasheed2014amw}.
    The weather data is available through the Met Norway Thredds Service~\cite{metthredds} using the Open-source Project for a Network Data Access Protocol (OPeNDAP).
    For automated online usage, the forecast can be accessed for selected regions and parameters in ASCII encoding through web requests through the URL constructed with
    
    \begin{lstlisting}[language=Python]
url = base_url + param_1 + "," + param_2 + ...
base\_url = "https://thredds.met.no/thredds/dodsC/mepslatest/meps\_lagged\_6\_h\_vc\_2\_5km\_"+ yyyy + mm + dd + "T" + hh + "Z.ncml.ascii?"
param_i = name + range_1 + range_2 + ...
range_i = "\%5B" + start + ":" + stepsize + ":" + end + "\%5D"
\end{lstlisting}
    
    Here $name$ refers to the name of the parameter. $start$, $stepsize$, and $end$ refer to the indices of each parameter, e.g. time, forecast ensemble member, height, grid-y-index, and grid-x-index. $yyyy$, $mm$, $dd$, and $hh$ refer to the forecast year, month, day, and hour respectively. Forecasts are performed every 6 hours at 00, 06, 12, and 18 o'clock.
    Note that the data is also available through NetCDF. Access through both options has been implemented in Python, but only ASCII strings have been implemented for direct access through Unity. For the onshore wind farm, data from a nested microscale simulation called SIMRA has been used from the same source. 

    While access from both Python and Unity has been implemented, access through Python is preferred and shown in the data integration framework as it allows the forecasts to be used in computations together with other online and offline data before forwarding it to the clients. 

    \subsection{OEM and operator data}
    While a static terrain and 3D model can be used for a standalone digital twin, they are not sufficient to describe changes to the asset. For all higher capability levels, the dynamic change of the asset needs to be described. Models can be used to approximate the asset response to metocean data but for a holistic description of the up-to-date asset state, additional real-time measurements from the asset are required.
    Often, data is being measured by multiple stakeholders.
    In wind energy and many other engineering-related industries, three types of stakeholders are the original equipment manufacturers, the operators, and potential third-party stakeholders that are contracted to measure for the manufacturer or operator, provide consulting services, or take measurements for other reasons.
    Even though all three stakeholders are interested in the same asset, they often have their own data accumulation systems and data interfaces, and frequently there are conflicts of interest, which can limit data sharing.
    
    Here, the resulting restrictions on proprietary data have made it impossible to access the manufacturer's and operator's data interfaces directly. Instead in these implementations, data sets have been provided as .csv files covering the data with operator access.
    Nonetheless, apart from proprietary restrictions, the software used for data accumulation by operator and manufacturer would allow establishing a real-time data stream to a Python environment through add-ons~\cite{KKwindsolutionscms}.
    Therefore, for the purpose of this work, the data is read as .csv but treated as real-time data from there onward.
    
    \subsection{Third party sensoring}
    In contrast to manufacturer and operator data streams, real-time access to the third-party data stream has been granted. The third-party measurements are performed by 4subsea. Access to the data is granted through their 4insight.io platform. The data is streamed to Python using their datareservoirio and fourinsight Python packages. Documentation can be found on \cite{4subsead, 4subseaf}. Since the data is proprietary, an authentication code is required to start the data stream, but afterward, the software can be used in real-time. For commercial use, a client secret can be used for continuous access.

    \subsection{Frontend-backend communication}
    \label{ssec:frontback}
    With all data connected to and processed in the computing component, the real-time data still needs to be received by the clients. The real-time communication between the computing component and the client is based on web requests. This choice is beneficial due to its low requirement for any client software, which makes it possible to easily deploy even on technologies with limited software support such as virtual reality headsets.

    On the computing component, a server has been set up in Python using the fastapi package, following this example:
    \begin{lstlisting}[language=Python]
# import packages
from fastapi import FastAPI
import nest_asyncio
import uvicorn
# instantiate the application
app = FastAPI()
# define a function without any parameters
@app.get("\myendpoint")
async def root():
    # get dictionary with data from a regular Python function
    data = get_data()
    # send the data online as RESTful
    return data
# circumvent Jupyter Notebook event loop
nest_asyncio.apply()
# run application
uvicorn.run(app, host="0.0.0.0", port=8000)
\end{lstlisting}
    In addition, the nest-asyncio and uvicorn packages are used. 
    First, the fastapi app is instantiated. Next, an endpoint is defined. The address extension in this example is $\backslash myendpoint$. Inside the corresponding function, any regular Python code can be executed, such that the data can be loaded as a dictionary through other functions. The function returns the dictionary so that by accessing the address of the endpoint, the data can be received. Different data can be accessed by adding additional endpoints.
    For applications running inside an event loop such as Jupyter Notebook or Google Colab, the nest\_asyncio.apply() command needs to be executed to allow nested use of asyncio.
    Finally, the app can be started using the uvicorn.run() function. The parameter host and port should be specified to have a fixed address inside a local network.
    Any other device can now access the endpoint using the web address composed as \textit{http://IP/port/endpoint} where the IP belongs to the computer/server hosting the Python script, the port is specified inside the run() function above, and the endpoint parameter determines which endpoint is accessed. To access the server from outside the local network, a VPN can be used to connect, which furthermore increases data security.
    
    Inside any Unity application, the HttpWebRequest class or UnityWebRequest module can be used to send a request and accept data.
    Here is an example adapted from~\cite{Unityusr}:
    \begin{lstlisting}[language=C++]
// include relevant packages
using Unity Engine;
using UnityEngine.Networking;
using System;
public class MyScript : MonoBehaviour{
void Start(){
    // start request in a coroutine so that the framerate is not impacted
    StartCoroutine(GetRequest("myuri"));
}
IEnumerator GetRequest(string uri){
    using (UnityWebRequest webRequest = UnityWebRequest.Get(uri)){
        // send the request
        yield return webRequest.SendWebRequest();
        // receive the data
        json = webRequest.downloadHandler.text;
        // transform the data back into the original 
        // datatype, replace mydatatype with the 
        // corresponding data type
        mydatatype data = JsonUtils.FromJson<mydatatype>(json);
        // do something with the data
        ProcessData(data);
    }
}
}
\end{lstlisting}
    When the script is activated, a coroutine is started to prevent the app from freezing while waiting on the response. "myuri" stands for the HTTP address of the endpoint. Inside the coroutine, the web request is sent, and the data is downloaded. Here data is received as a string, but it can be transformed back into the original data format using JsonUtils.

\section{Results and Discussion}
\begin{figure}[h]
    \centering
    \includegraphics[width=\linewidth]{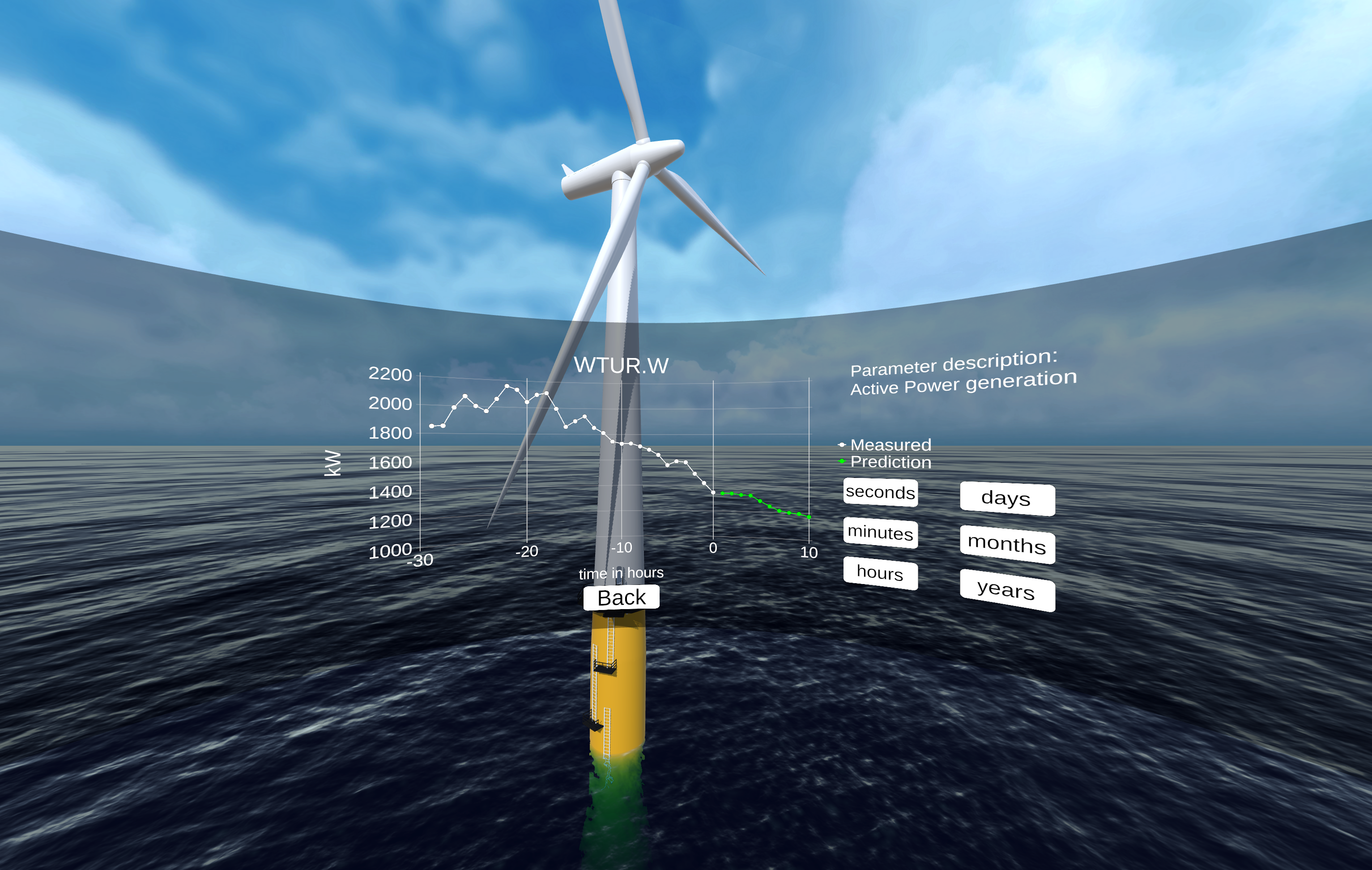}
    \caption{Data integrated into the floating offshore wind turbine}
    \label{fig:zefyros}
\end{figure}

\begin{figure}[h]
    \centering
    \includegraphics[width=\linewidth]{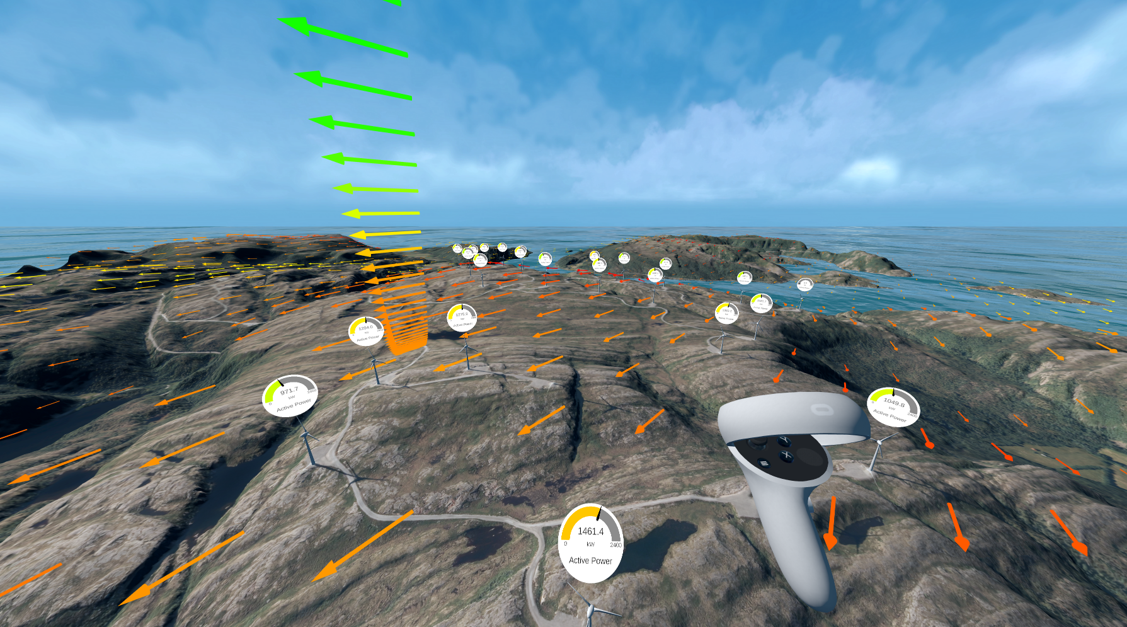}
    \caption{Data integrated into the onshore wind farm digital twin 
    }
    \label{fig:bessaker}
\end{figure}

The integration framework has been tested through successful implementation for two real user cases, namely the onshore wind farm Bessakerfjellet and the floating offshore wind turbine Zefyros. 
In Figure~\ref{fig:zefyros}, the digital twin of Zefyros is shown as the streamed data from one of the sensors is depicted graphically.
In Figure~\ref{fig:bessaker}, the implemented digital twin of Bessakerfjellet is shown through a virtual reality headset. The streamed meteorological data is shown through vectors, while turbine measurements are visualized through gauges, and the underlying terrain can be used to interpret wind flow and identify turbine access roads.

While this work presents an integration framework for digital twins with real-time data streams, there is more data being measured at wind turbines where access was restricted. As even operators themselves often do not have access to all data, more data sharing is essential to enable digital twins in wind energy and similar areas on an industrial scale. Additionally, the asset data here has been categorized by data owners since they represent the access points, which are of importance to building the data streams. However, for an industrial application, the data stream needs to be broken down into single-sensor measurements, since their designations need to be standardized. Nonetheless, this does not diminish the value of the current work, as access to those data components will still happen through the data owner's access point.
Finally, the fragmentation of the modeling components has been out of the scope of this work, since the required modules and languages highly depend on the asset and the purpose of the digital twin. It should be noted, that still the modules can still typically be connected to Python, which does not limit the value of the integration framework.

Despite these limitations of the work presented in this article, it contributes towards easier implementation of digital twin environments, real-time data streams, and utilization of virtual reality devices for data visualization in real-time.

\section{Conclusion}
In this work, a multi-source, multi-client data integration framework for digital twins was presented and explained with a focus on wind energy, but applicable to many other areas.
The data framework can accommodate multiple data sources and multiple clients across different devices.
Implementation and sources of environmental quantities have been explained, both static for terrain height, ocean depth, and land cover, as well as in real-time for weather forecasts.
Furthermore, it has been demonstrated how a server can be set up through a Python script to transmit data and receive it on different devices, including virtual reality headsets, where it can be used for improved data visualization.
Additionally, the feasibility of real-time data transfer from a floating wind turbine has been shown.
Finally, the strength and shortcomings of the presented data integration framework are being discussed in the context of digital twins and wind energy in general, and avenues for further research has been pointed out.

\section*{Acknowledgments}
This publication has been prepared as part of NorthWind (Norwegian Research Centre on Wind Energy) co-financed by the Research Council of Norway (project code 321954), industry, and research partners. Read more at \url{www.northwindresearch.no}.
We thank Aneo, SINTEF Energy, 4subsea, and Sustainable Energy Catapult Center for providing wind turbine data for the project.

\bibliographystyle{AR}
\bibliography{references}
\vspace{12pt}

\end{document}